\documentclass[prc,showpacs,twocolumn,amsmath,amssymb]{revtex4}

\usepackage{graphicx}
\usepackage{dcolumn}
\usepackage{bm}

\newcommand{\bra}[1]{\langle {#1} |}
\newcommand{\ket}[1]{| {#1} \rangle}
\newcommand{\inproduct}[2]{\langle #1 | #2 \rangle}

\begin{document}


\title{Configuration mixing calculation for complete low-lying spectra
with the mean-field Hamiltonian}

\author{Satoshi Shinohara}
\affiliation{Institute of Physics, University of Tsukuba, Tsukuba 305-8571, Japan}
\author{Hirofumi Ohta}%
\affiliation{Sumitomo Chemical Co. Ltd, Tsukuba 300-3294, Japan}
\author{Takashi Nakatsukasa}
\author{Kazuhiro Yabana}
\affiliation{Institute of Physics, University of Tsukuba, Tsukuba 305-8571, Japan}
\affiliation{Center for Computational Sciences, University of Tsukuba, 305-8571, Japan}

\date{\today}

\begin{abstract}
We propose a new theoretical approach to
ground and low-energy excited states of nuclei extending the
nuclear mean-field theory. It consists of three steps:
stochastic preparation of many Slater determinants,
the parity and angular momentum projection,
and diagonalization of the generalized eigenvalue problems.
The Slater determinants are constructed
in the three-dimensional Cartesian coordinate representation
capable of describing arbitrary shape of nuclei.
We examine feasibility and usefulness 
of the method by applying the method with
the BKN interaction to light 
$4N$-nuclei, $^{12}$C, $^{16}$O, and $^{20}$Ne.
We discuss difficulties of keeping linear independence
for basis states projected on good parity and angular momentum
and present a possible prescription.
\end{abstract}

\pacs{21.60.-n, 21.30.Fe}
\maketitle

\section{\label{sec:introduction}Introduction}

One of the goals in the microscopic nuclear many-body theory is
the ab-initio nuclear structure calculation starting with a fixed
Hamiltonian.
Indeed, recent Green's function Monte-Carlo calculation with a bare
nucleon-nucleon (NN) interaction is a milestone
in this direction \cite{GFMC1,GFMC2,GFMC3,GFMC4,GFMC5}.
However, these ab-initio calculations are still limited to
nuclei with mass number less than around twelve and to the lowest
energy state for each parity and angular momentum.
The no-core shell model \cite{NCSM}
utilizes unitary transformation to accommodate the short-range
correlations.
These introduce effective NN interactions (without phenomenological
adjustments), however their applications are also limited to
light nuclei near the closed configurations.
Systematic description of ground and excited states in nuclei
in a wide mass region requires
development of a new computational approach.

One of difficulties of the nuclear many-body problem
is due to the strong short-range correlation.
This forbids a naive mean-field approach using the bare NN interaction.
In order to overcome this difficulty,
effective NN interactions have been extensively developed in
history of the nuclear theory \cite{RS}.
Namely, the short-range behavior of two-body correlation
is effectively renormalized in the interaction.
Then, nuclear many-body wave functions should take account of
long-range correlations only.
Most of microscopic nuclear structure models adopt
the effective interactions;
e.g., nuclear mean-field models \cite{SHF1,SHF2,Gogny,RMF},
shell models \cite{shellmodel1,shellmodel2}, cluster models \cite{cluster}, etc.
The success of these models indicates that
a variety of low-lying modes of excitation are governed by nothing but the
long-range correlations.
In the present study,
we aim for developing a new method to treat the whole correlation of long range
beyond the mean field,
utilizing the effective interaction for the mean-field models.

To test our theory, we will apply the method to light $N=Z$ even-even nuclei.
There are a variety of nuclear models for light nuclei.
Yet, the existing models are not satisfactory in some respects.
The shell model nicely describes spectroscopic properties of light nuclei.
However, since the model space is truncated to a specific 
shell, states very different from the ground state,
which require a wider space, cannot be described adequately. 
A classic example is the second $J^\pi=0^+$ state in $^{16}$O.
Although this is the lowest-lying excited state of this doubly closed-shell 
nucleus, the shell model fails.
The state has been successfully described by the cluster model.
The nuclear cluster models have provided a fruitful description of 
many light nuclei, especially for those states lying close to the 
threshold.
The antisymmetrized molecular dynamics (AMD),
which was first utilized for study of heavy-ion collision \cite{AMD},
is an extension of the microscopic cluster model
successful to describe shell-model-like states as well
\cite{AMD_VAAMP1,AMD_VAAMP2}.
However, the model space of the AMD
is a superposition of small number of
Slater determinants whose orbitals are restricted to a Gaussian form.
These models, the shell model and the AMD,
are applicable to either light nuclei or those near
the closed configuration.
In contrast, the nuclear mean-field theory has been successful 
to describe nuclei of a wide mass region with a few parameters associated
with the (density-dependent) effective nuclear force.
The theory provides a reasonable description with a single Slater determinant
for total binding energy, radius, and deformation of ground states.
However, the superposition of multiple Slater is often required.
For example, when the mean-field solution violates symmetries of
the original Hamiltonian, one should superpose many Slater determinants to
restore the symmetry (``projection'') \cite{RS}.
The symmetry restoration is crucial for many properties of light nuclei.
Recently, we have developed a method of
the parity (angular momentum) projection before (after)
variation, and have shown that the mean-field model
is capable of describing some typical cluster structures in light nuclei
\cite{PPSHF}.
In this paper,
we intend to further extend the method to a kind of {\it complete}
calculation of the long-range correlations, to
obtain energies and wave functions
for the ground and low-lying excited states starting from a nuclear
mean-field Hamiltonian.

Our method has some resemblance to the generator coordinate method
(GCM) \cite{HW,RS} and the Mote Carlo shell model (MCSM) \cite{MCSM}.
In the GCM, the generator coordinate is adopted 
{\it a priori}, under a certain physical intuition, to describe
specific long-range correlations;
e.g., quadrupole and octupole correlation.
In most practical calculations,
the coordinate is limited to one dimension.
Our method stochastically take into account all the important correlations.
In the MCSM, basis states are stochastically generated and selected,
then the diagonalization of the Hamiltonian is performed in the space
spanned by those states.
This concept is very similar to ours, but
we use the mean-field-model Hamiltonian and
our model space is much wider than that of the shell model,

The paper is organized as follows.
In Sec.~\ref{sec:formalism}, we present the outline of our method.
Its details are described in the following sections;
Selection of Slater determinants
and the parity and angular momentum projection are shown in
Secs.~\ref{sec:S-det} and \ref{sec:projection}, respectively.
In Sec.~\ref{sec:mixing}, we discuss
configuration mixing
calculation and how to avoid numerical instability caused by the
overcompleteness of a selected basis and numerical errors.
Then, we test the accuracy of our approach 
by taking $^{16}{\rm O}$ as an example.
In Sec.~\ref{sec:C_Ne}, 
we compare numerical results of $^{12}{\rm C}$ and 
$^{20}{\rm Ne}$ to experimental data.
The summary is given in Sec.~\ref{sec:summary}.

\section{\label{sec:formalism}Formalism}

In this section, we present the outline of our method to illustrate
its essence.
Roughly speaking,
our method consists of three steps;
\renewcommand{\labelenumi}{(\theenumi)}
\begin{enumerate}
\item Generation and selection of Slater determinants (S-det's)
      important for ground and low-lying excited states.
\item Parity and angular momentum projection.
\item Configuration mixing (diagonalization of the Hamiltonian).
\end{enumerate}
Each of these steps is not as straightforward as it first looks.
For the step (1), since the short-range correlation is renormalized in
the effective interaction,
we should be careful not to adopt a S-det involving components
with very high momentum.
For the step (2), since there is no symmetry restriction on the wave function,
we need to carry out the projection with respect to
the full three-dimensional Euler angles.
The diagonalization in (3) is cursed by the well-known overcompleteness
problem of non-orthogonal basis and also by
accumulated numerical errors in the step (2).
We will present prescriptions to overcome these problems in
Secs.~\ref{sec:S-det}, \ref{sec:projection}, and \ref{sec:mixing}, respectively.

Following the prescription given in Sec.~\ref{sec:S-det},
many S-det's are stochastically generated,
then those important for low-energy excitations are selected.
Single-particle wave functions in each S-det are represented
in the three-dimensional (3D) Cartesian coordinate space without any
symmetry restriction.
The selected S-det's form a basis set,
$\{ \ket{\Phi_{n}}; n=1,\cdots,N \}$.
We then make
parity and angular momentum projection for each S-det.
The Hamiltonian and the norm kernels for the fixed parity ($\pi=\pm$)
and angular momentum $(J,K)$, where $K$ indicates its body-fixed component,
are given by
\begin{equation}
\left\{ \begin{array}{c}
H_{nK,n'K'}^{J(\pm)} \\
N_{nK,n'K'}^{J(\pm)} \end{array} \right\}
=
\langle \Phi_n \vert \left\{ \begin{array}{c}
H \\
1 \end{array} \right\}
{\hat P}^\pm {\hat P}^J_{KK'} \vert \Phi_{n'} \rangle .
\end{equation}
Here, $\hat H$ is the many-body Hamiltonian with effective interaction, 
${\hat P}^\pm$ is the parity projection operator, and
${\hat P}^J_{KK'}$ is the angular momentum projection operator.
Finally, we solve the following generalized eigenvalue problem,
\begin{equation}
\sum_{n'K'} (H^{J(\pm)}_{nK,n'K'} -E^{J(\pm)} N^{J(\pm)}_{nK,n'K'} )
  g_{n'K'} = 0 .
\label{eq:GEE}
\end{equation}
If the space spanned by the set of the S-det's,
$\{ \ket{\Phi_{n}}; n=1,\cdots,N \}$, is approximately complete
for the long-range correlation,
we should obtain a convergent solution for the ground
and the low-lying excited states.
Note that this is merely an outline of the method.
As a matter of fact, in order to avoid the zero eigenvalues of
the norm kernels,
we will screen the selected basis states and modify Eqs. (\ref{eq:GEE}).
This prescription will be given in Sec.~\ref{sec:mixing}.

To demonstrate applicability of our method, we show
numerical calculations employing the simplified mean-field
Hamiltonian, the so-called BKN force \cite{BKN}.
The BKN force consists of two-body plus three-body forces.
The two-body force consists of those of zero-range part ($t_0$),
finite-range Yukawa part ($V_\text{Y}$), and the Coulomb part ($V_\text{C}$). 
The three-body term is a zero-range force ($t_3$).
The Hamiltonian is given by
\begin{equation}
\label{BKN}
H = \sum_{i=1}^A\left(-\frac{\hbar^2}{2m}\right)\nabla_i^2+ 
+ \frac{1}{2} \sum_{ij} V^{(2)}(ij)
 + \frac{1}{6} \sum_{ijk} V^{(3)}(ijk), 
\end{equation}
where
\begin{eqnarray*}
&&V^{(2)}(ij)=t_0\delta(\vec{r}_i-\vec{r}_j)
            +V_\text{Y}(\vec{r}_i-\vec{r}_j)
            +V_\text{C}(\vec{r}_i-\vec{r}_j) ,\\
&&V^{(3)}(ijk)=t_3\delta(\vec{r}_i-\vec{r}_j)\delta(\vec{r}_j-\vec{r}_k) .
\end{eqnarray*}
The BKN interaction
assumes that all nucleons have a charge $e/2$, and
four nucleons with different spin and isospin occupy the same
spatial orbital.
To express the orbital wave functions, we employ 
a grid representation discretizing 3D Cartesian coordinate.
Grid points inside a sphere of 8 fm are adopted with grid spacing of 0.8 fm.

\section{\label{sec:S-det}Preparation of basis}
%
\begin{figure}[tb]
\begin{minipage}{0.4\textwidth}
\includegraphics[width=\textwidth]{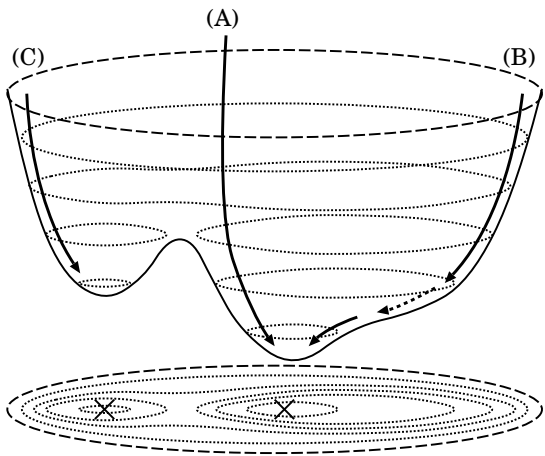}
\end{minipage}
\caption{\label{fig:ITSM1}
Schematic picture of the energy surface.
Two crosses represents minima of the energy surface.
Three paths, (A), (B) and (C),
show imaginary-time trajectories starting from different initial S-det's.
The dotted arrow of (B) indicates the trajectory (B) passes through
a shoulder state.
}
\end{figure}

\subsection{Generation and selection of S-det's}

The first step is to prepare a set of S-det's 
that span the space necessary for low-lying states with
the full long-range correlations.
We make use of the imaginary-time method starting from initial
configurations which are generated stochastically.
The imaginary-time method is often used to obtain
self-consistent solutions.
Instead, we utilize it for generating many kinds of
low-energy collective surfaces.
We pick up S-det's on the way to the self-consistent solutions
before reaching minima, and employ them in the 
configuration mixing calculation.

Figure~\ref{fig:ITSM1} shows a schematic picture of the imaginary-time 
calculations starting from different initial configurations.
The imaginary-time iteration has a property suitable for
generating the basis to calculate the long-range correlations.
It quickly removes high-energy components of the wave function
in a early stage of the iteration.
The S-det is expected to rapidly fall onto
a potential energy surface important for low-energy modes of excitation.
This is the very property we want, because we should exclude
S-det's which take account of the short-range correlation in the Hamiltonian.
Therefore, we simply dispose all the S-det's generated
in the first few hundred iterations of the imaginary-time evolution, 
then select S-det's after the rate of energy decrease becomes relatively slow.

A series of S-det's generated with the imaginary-time calculation
starting from an arbitrary initial state
eventually converge to a self-consistent solution;
either the Hartree-Fock ground state (path (A) and (B) in Fig.~\ref{fig:ITSM1})
or local minima solutions (path (C) in Fig.~\ref{fig:ITSM1}).
During the iterations, 
it sometimes happens that the configuration changes very slowly and 
the state stays almost unchanged for a long period of the iterations
(a part presented by the dotted arrow of path (B)).
This is called a shoulder state.
Although these shoulder states are 
not self-consistent solutions, they may play an important role for
the low-lying excitation spectra and the ground-state correlation.

We repeat the imaginary-time iteration many times starting from different
initial configurations.
We construct the initial S-det's by a stochastic procedure:
The single-particle 
orbitals of the initial S-det are in a
Gaussian form whose centers are randomly chosen.
After generating large number of imaginary-time trajectories,
we may expect that those S-det's span the complete space
for calculation of the long-range correlations.

\begin{figure}[tb]
\begin{minipage}{0.48\textwidth}
\includegraphics[width=\textwidth]{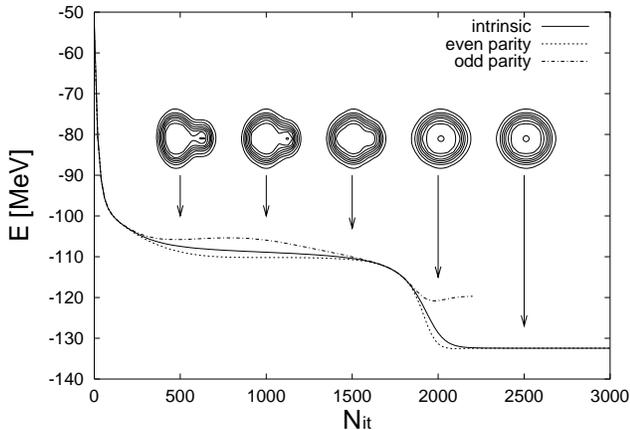}
\end{minipage}
\caption{\label{fig:ITSM2}
An example of the imaginary-time evolution in 
$^{16}{\rm O}$ started from a randomly generated S-det.
Solid line indicates energy expectation value
of the S-det, $\ket{\Phi(N_\text{it})}$,
as a function of iteration number, $N_\text{it}$.
Snapshots of the density distribution 
at every 500 iterations are shown.
The dashed and the dash-dotted 
line indicate energy of even and odd parity component of the S-det,
respectively.
The imaginary-time step of 
$\Delta \tau = 0.001$ $\hbar/$MeV is adopted in the calculation.
}
\end{figure}

Figure~\ref{fig:ITSM2} is an example of the actual imaginary-time
calculations for $^{16}$O, showing the energy expectation value,
$E(N_\text{it})=\bra{\Phi(N_\text{it})}H\ket{\Phi(N_\text{it})}$, as a
function of the iteration number, $N_\text{it}$.
The path is similar to (B) in Fig.~\ref{fig:ITSM1},
passing through a shoulder state.
In $N_\text{it}<100$, the energy expectation
value decreases very rapidly.
From $N_\text{it}=200$ to 1500, the
energy decreases very slowly, corresponding to a shoulder state. 
We have found that this shoulder state
corresponds to the cluster structure of $^{12}$C+$\alpha$,
which is considered as a dominant component of the first excited 
state of $^{16}$O in the cluster model studies. 
The dashed and dash-dotted curves are the energy 
expectation value after parity projection,
$E^{(\pm)}(N_\text{it})=
\bra{\Phi^{(\pm)}(N_\text{it})}H\ket{\Phi^{(\pm)}(N_\text{it})} /
\inproduct{\Phi^{(\pm)}(N_\text{it})}{\Phi^{(\pm)}(N_\text{it})}$,
where $\ket{\Phi^{(\pm)}(N_\text{it})}=P^\pm\ket{\Phi(N_\text{it})}$.

\subsection{Selection of S-det's}

During the imaginary-time iterations of $N_{\rm total}$ steps,
S-det's for every $N_\text{s}$ iterations are taken as 
candidates of the basis states.
Thus, the S-det's at
$N_\text{it}^\text{c}=N_\text{s},2N_\text{s},\cdots,k_\text{n}N_\text{s}$
are nominated first.
The number of S-det's taken from a single path is
$k_\text{n}=N_{\rm total}/N_{\rm s}$.
The typical numbers are $N_{\rm s}=50$ and $N_\text{total}=2000$,
leading to $k_{\rm n}=40$.
However, we cannot include all these S-det's in the basis set of the
configuration mixing calculation,
because too many S-det's lead to a numerical instability caused by
the overcompleteness.
Thus, we need to reduce their number.
Here, we impose two additional constraints on those candidates:
\renewcommand{\theenumi}{\alph{enumi}}
\begin{enumerate}
\item \label{a}
$E(N_\text{it}^\text{c}) <E_\text{HF}+30$ MeV.
\item \label{b}
Overlap between any pair of selected S-det's must be less than 0.7
(see below for details).
\end{enumerate}
The condition (\ref{a}) means that the energy expectation value of
each S-det, $E(N_\text{it}^\text{c})=
 \bra{\Phi(N_\text{it}^\text{c})}H\ket{\Phi(N_\text{it}^\text{c})}$,
should not be so large because we are interested in low-lying states and
the long-range correlations only.
In the present work, we adopt the cut-off energy as 30 MeV above the
Hartree-Fock ground-state energy.

The second condition is directly related to the linear independence among 
the S-det's.
In order to determine whether to include a candidate in the basis set,
we examine the overlaps between the new S-det (candidate) and all the 
S-det's which have been already included in the basis set.
If the maximum of the absolute values of the overlap
is less than a certain value,
we add the candidate to the set of basis states.
Since we make parity and angular momentum projection later,
it is desirable to check this condition for projected wave functions.
However, it costs too much to achieve angular momentum projection
at this stage. 
Instead, we examine the overlap for different configurations produced by
the interchange and the inversion of the Cartesian axes.
These transformations correspond to 24 choices of the coordinate system,
and are easily realized in the 3D Cartesian coordinate representation.
The condition (\ref{b})
for adding a new S-det $\ket{\Phi(N_\text{it}^\text{c})}$
to the selected basis set $\{ \ket{\Phi_n}; n=1,\cdots, M \}$
is expressed by
\begin{eqnarray}
\frac{|\bra{\Phi(N_\text{it}^{\rm c})} P^{\pm} {\hat R}^i\ket{\Phi_n}|}
{|\bra{\Phi(N_\text{it}^{\rm c})} P^{\pm} \ket{\Phi(N_\text{it}^{\rm c})}
 \bra{\Phi_n} P^{\pm} \ket{\Phi_n}|^{\frac{1}{2}}}
<0.7 , \nonumber\\
\quad\quad\quad \mbox{ for } n=1,\cdots, M ,
\label{eq:condition}
\end{eqnarray}
where $\hat{R}^i, i=1 \cdots 24$ are special rotations and inversions 
corresponding to permutation of the axes $(x,y,z)$.

In practice, the Hartree-Fock state $\ket{\Phi_\text{HF}}$ is
always selected first.
Then, we start the examination of constraints (\ref{a}) and (\ref{b}) for
generated S-det's $\ket{\Phi(N_\text{it}^\text{c})}$ in the
ascending order of the energy expectation value.
Here, we employ 
not the energy expectation values with respect to the S-det,
but those with respect to the states projected onto negative parity,
$P^- \ket{\Phi(N_\text{it}^\text{c})}$.
This makes it easier to select exotic deformations
which often appear in the negative-parity excited states.
For the case in Fig.~\ref{fig:ITSM2}, states
at $N_\text{it}\approx 2000$ are examined first.
Since those around $N_\text{it}\approx 500$
also show local minima in the energy surface of
negative parity, they are also nominated with high priority. 

The number of S-det's satisfying the criteria (\ref{a}) and (\ref{b}) are
typically from zero to three in a single trajectory of the
imaginary-time evolution.
Apparently, larger the number of selected S-det's, $M$,
more difficult it becomes to find the $(M+1)$-th S-det to
satisfy the condition (\ref{b}).
In actual calculations,
about 100 imaginary-time trajectories will be repeatedly generated to
obtain about fifty S-det's which satisfy these conditions.

\section{\label{sec:projection}Parity and angular momentum projection}

\subsection{Projection with respect to 3D Euler angles}

For each S-det in a set $\{ \ket{\Phi_n}; n=1,2,\cdots,N$ \}
which are generated and selected in Sec.~\ref{sec:S-det}.
we perform the parity and angular momentum projection.
Since we construct wave functions employing the 3D Cartesian
coordinate representation without any restrictions on the
spatial symmetries, the full three-dimensional angular momentum projection 
is necessary.
The three Euler angles,
$(\alpha, \beta, \gamma)$, are abbreviated as $\Omega$.
The angular momentum projection operator $P^J_{MK}$ is given as usual by
\begin{equation}
\label{eq:AMP}
P^{J}_{MK}=
\frac{2J+1}{8\pi^{2}}\int \! d\Omega D^{J*}_{MK}(\Omega){\hat R}(\Omega),
\end{equation}
where $D^J_{MK}(\Omega)$ is the Wigner's $D$-function and
$\hat R(\Omega)$ is the rotation operator.

Quantities necessary for solving a generalized eigenvalue equation
in Sec.~\ref{sec:mixing} are matrix elements of
the norm and Hamiltonian kernels.
For two S-det's, $\ket{\Phi_n}$ and $\ket{\Phi_{n'}}$, we
calculate matrix elements for parity and angular momentum projected
wave functions.
\begin{widetext}
\begin{eqnarray}
\left\{\!\!
\begin{array}{c}
N^{J(\pm)}_{nK,n'K'}\\
H^{J(\pm)}_{nK,n'K'}
\end{array}
\!\!\right\}
&\equiv&
\bra{\Phi_n}
P^{\pm\dagger}P^{J\dagger}_{MK}
\left\{\!\!
\begin{array}{c}
1\\
{\hat H}
\end{array}
\!\!\right\}
P^{J}_{MK'}P^{\pm}\ket{\Phi_{n'}}
=\bra{\Phi_n}
\left\{\!\!
\begin{array}{c}
1\\
{\hat H}
\end{array}
\!\!\right\}
P^{J}_{KK'}P^{\pm}\ket{\Phi_{n'}}\nonumber\\
&=&\frac{2J+1}{16\pi^{2}}\int\! d\Omega D^{J*}_{KK'}(\Omega)
\times\bra{\Phi_n}
e^{-i\alpha{\hat J}_{z}}
\left\{\!\!
\begin{array}{c}
1\\
{\hat H}
\end{array}
\!\!\right\}
(1\pm{\hat P})
e^{-i\beta{\hat J}_{y}}
e^{-i\gamma{\hat J}_{z}}
\ket{\Phi_{n'}} ,
\label{eq:ME}
\end{eqnarray}
\end{widetext}
where $K$ and $K'$ specifies quantum numbers for body-fixed component 
of the angular momentum operator.
Since the rotation operator commutes with the Hamiltonian, 
we apply rotations of angles $\beta$ and $\gamma$
to the ket vector $\ket{\Phi_{n'}}$
and the rotation of $\alpha$ to the bra vector $\bra{\Phi_n}$
in the practical calculations.
The parity projection operator is given by 
$P^{\pm}=(1/2)(1\pm{\hat P})$, where ${\hat P}$ is the space inversion operator.

The parity transformation and rotation of the S-det,
$\ket{\Phi}=\det \{ \ket{\phi_k} \}/\sqrt{A!}$,
are achieved by the corresponding transformation 
of their single-particle orbitals, $\ket{\phi_k}$, $k=1,\cdots, A$.
The rotation of finite angle is realized by successive rotations of a 
small angle.
For instance, for a rotation of angle $\alpha$ around $z$-axis,
\begin{equation}
\ket{\phi_k^\alpha} \equiv
e^{-i\alpha \hat{j}_z} \ket{\phi_k} =
 \left( e^{-i \Delta\alpha \hat{j}_z}\right)^{N_\text{div}} \ket{\phi_k} ,
 \quad \Delta\alpha=\alpha/N_\text{div} .
\end{equation}
Each small-angle rotation is performed
using the Taylor expansion of the rotation operator;
\begin{eqnarray}
\ket{\phi^{\alpha+\Delta\alpha}_k}
&=&e^{-i\Delta\alpha\hat{j}_{z}}\ket{\phi^{\alpha}_k}\nonumber\\
&\approx&\sum^{N_\text{max}}_{k=0}\frac{(-i\Delta\alpha \hat{j}_z)^{k}}{k!}
\ket{\phi^{\alpha}_k},
\label{eq:Taylor}
\end{eqnarray}
where $N_\text{max}=4$ gives an accurate result. 
We usually employ $\Delta \alpha= 2\pi/360$.

The integrand of Eq. (\ref{eq:ME}) is
the overlap/Hamiltonian matrix element between
two different S-det's,
$e^{-i\alpha \hat{J}_z}\ket{\Phi_n}$ and
$ (\hat{P}) e^{-i\beta \hat{J}_y} e^{-i\gamma \hat{J}_z}\ket{\Phi_{n'}}$.
These matrix elements are simply expressed in terms
of the interstate density matrix defined in Appendix.

\subsection{Numerical details of the projection}

We here discuss numerical accuracy of the 3D angular momentum projection.
Numerical error in the matrix elements of Eq.~(\ref{eq:ME})
may cause a serious trouble when we solve the generalized eigenvalue problem
of Eq.~(\ref{eq:GEE}).
For instance, the norm matrix, $N_{nK,n'K'}^{J\pm}$,
should be positive definite,
however, in practice,
calculated norm matrix suffers from many negative eigenvalues
though their absolute values are small.

The finite difference approximation for the angular momentum
and the finite-order expansion 
for the rotation operator in Eq.~(\ref{eq:Taylor}) are
good approximation.
We have examined the identity of the single-particle orbitals
before and after rotating over $2\pi$.
The overlap between these two single-particle wave functions
is very close to unity, with error less than $10^{-4}$.
Therefore, the error in each rotated wave function is relatively small.
However, it seems that these numerical errors are accumulated 
during the 3D integration over Euler angles,
$0\leq \alpha<2\pi$, $0\leq \beta<\pi$, and $0\leq \gamma<2\pi$.

\begin{figure*}[tb]
\begin{minipage}{0.49\textwidth}
\includegraphics[width=\textwidth]{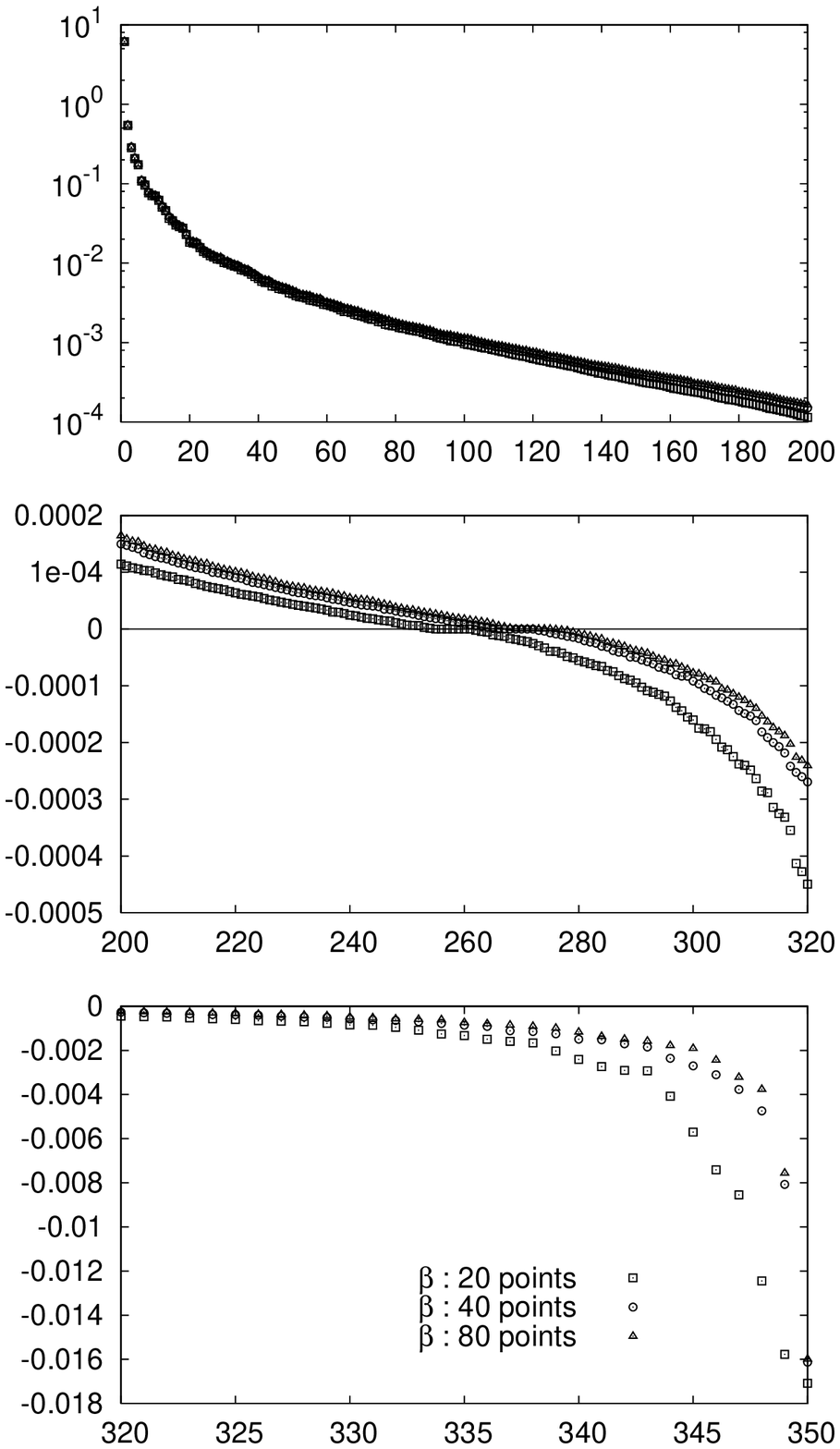}
\end{minipage}
\begin{minipage}{0.49\textwidth}
\includegraphics[width=\textwidth]{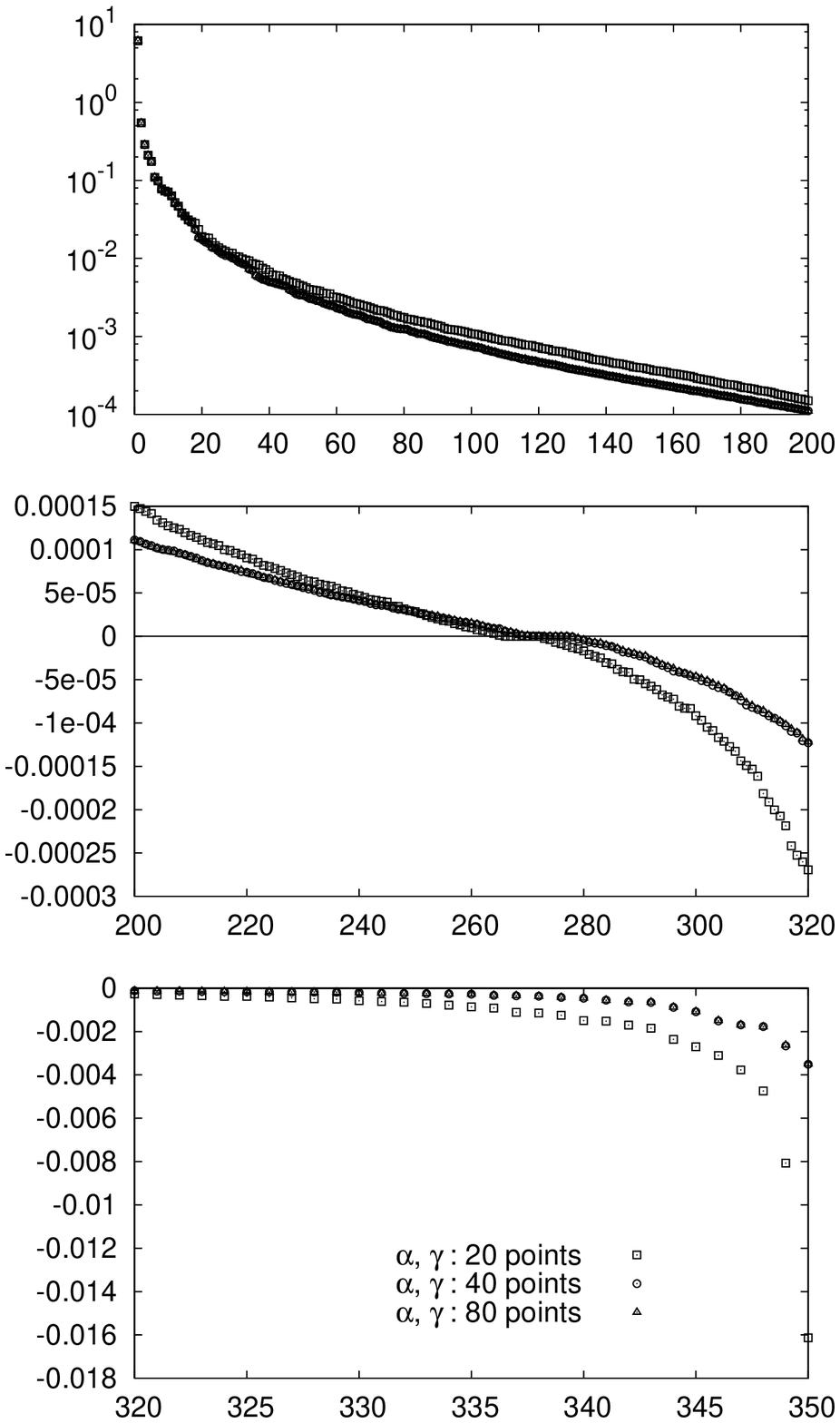}
\end{minipage}
\caption{\label{fig:Euler}
Eigenvalues of norm matrix for $J^\pi=3^-$ in $^{16}$O.
The horizontal line indicates the sequential number according to
the magnitude of the eigenvalues.
The top, middle and bottom panels show those of
No. 1 to 200, 200 to 320, and 320 to 350, respectively.
The norm matrix is calculated with various
discretization on Euler angles.
In the left panels,
$\beta$ is discretized into 20 (squares), 40 (circles), and 80 (triangles)
points, while $\alpha$ and $\gamma$ are discretized into 20 points.
In the right panels, $\alpha$ and $\gamma$ are discretized
into 10 (squares), 20 (circles), and 40 (triangles) points,
while $\beta$ is into 20 points.
}
\end{figure*}

The numerical integration is carried out using the trapezoidal rule
with the finite-number discretization.
Figure~\ref{fig:Euler} shows the norm eigenvalues calculated
for $J^{\pi}=3^-$ in $^{16}$O.
Fifty Slater determinants 
are generated in the procedure explained in Sec.~\ref{sec:S-det},
thus the dimension of the norm matrix is 350 ($=7 \times 50$ where
7 is the number of different $K$ quantum numbers).
The eigenvalues of 
norm matrix $N^{3(-)}_{nK,n'K'}$ are plotted in descending order. 
The left three panels show the eigenvalues
when the number of grid points of the angle $\beta$ is varied from 20 to 80 
while those for $\alpha$ and $\gamma$ are fixed at 20.
For the right panels, we vary the number for
$\alpha$ and $\gamma$ from 10 to 40, being fixed at 20 for $\beta$.
The norm eigenvalues are converged for $\alpha$ 
and $\gamma$ if we adopt 15 or more grid points for their discretization.
In contrast,
the convergence with respect to $\beta$ is rather slow and
we still have about 70 negative eigenvalues with 80 grid points.
Apparently, it is desirable to have larger number of grid points
for the discretization of $\beta$ angle.
However, we must make a compromise and sacrifice some accuracy,
because the calculation for the Hamiltonian kernels
in Eq.~(\ref{eq:ME}) are very demanding in the 3D rotation.
In the present work,
we employ discretization of 15 grid points for $\alpha$ and $\gamma$ 
and 20 points for $\beta$.
Even with this discretization,
we need to evaluate 4,500 matrix elements for each pair
of the S-det's.

A price of sacrificing the accuracy is a complication
of solving the eigenvalue problem,
due to appearance of negative norm.
In Sec~\ref{sec:mixing}, we will explain how to cope
with this difficulty.

\section{\label{sec:mixing}Configuration mixing; energy spectra of $^{16}$O}

\subsection{Zero- and negative-norm problem}

Although we collect a set of linearly independent S-det's
in the procedure explained in Sec.~\ref{sec:S-det},
the linear independence is often lost after the
parity and angular momentum projection.
This will lead to number of eigenvalues of the norm matrix close to zero.
Moreover, the numerical error in the angular momentum projection 
even produces the negative eigenvalues.
In order to solve the eigenvalue equation (\ref{eq:GEE}),
we must remove states which cause the zero and negative
eigenvalues.
In this section, we give a possible prescription for this.

For each angular momentum state $J$, we reduce the dimension
of the configuration space as follows.
We first diagonalize the norm matrix in different $K$ states for each S-det.
This is the diagonalization of $(2J+1)\times (2J+1)$ matrix,
\begin{eqnarray}
\sum_{K'=-J}^J N^{J(\pm)}_{nK,nK'}v^{n\nu}_{K'}=
\lambda_{n\nu}^{J(\pm)}v^{n\nu}_{K}.
\end{eqnarray}
Here the eigenstates are labeled by $\nu$.
After the $K$-diagonalization, the basis 
is specified by the label $\nu$, instead of $K$.
The collected basis states are denoted as
$\{ \ket{\Psi_{n\nu}^{J(\pm)}}\}_{n,\nu,J}$;
\begin{equation}
\ket{\Psi_{n\nu}^{J(\pm)}}_M=\sum_{K}v^{n\nu}_{K}P^{J}_{MK}P^{\pm}\ket{\Phi_n} ,
\quad M=-J,\cdots,J .
\label{eq:base1}
\end{equation}
The magnetic quantum number in the laboratory frame, $M$, is a trivial
conserved quantity simply giving the $(2J+1)$-fold degeneracy for each state,
thus omitted hereafter.
At this stage, we exclude
states whose eigenvalues $\lambda_{n\nu}^{J(\pm)}$
are smaller than $10^{-2}$.
A small eigenvalue will appear when, for example, the S-det
is an approximate eigenstate of a specific $K$ quantum number.
In the case of $J^{\pi}=3^-$ of $^{16}$O, we have 350 configurations
constructed from fifty S-det's,
among which about one hundred
configurations have $\lambda_{n\nu}^{J(\pm)}$ less than $10^{-2}$ and
are discarded.

Although the states in $\{ \ket{\Psi_{n\nu}^{J(\pm)}}\}_{n,\nu,J}$
are truncated according to the norm eigenvalues,
in order to avoid numerical instability,
we need to reduce the number furthermore.
For this purpose, we consider a following ``normalized'' norm matrix,
\begin{equation}
{\tilde N}^{J(\pm)}_{n\nu, n'\nu'}\equiv
\frac{N^{J(\pm)}_{n\nu, n'\nu'}}
    {\left(N^{J(\pm)}_{n\nu, n\nu}\right)^{1/2}
     \left(N^{J(\pm)}_{n'\nu',n'\nu'}\right)^{1/2}},
\end{equation}
which is constructed so as to make the diagonal elements
${\tilde N}^{J(\pm)}_{n\nu n\nu}$ equal to unity.
We make a further selection according to the magnitude of eigenvalues of this
matrix, $\tilde{\lambda}_i$, obtained by solving
\begin{equation}
\sum_{n'\nu'}{\tilde N}^{J(\pm)}_{n\nu,n'\nu'}u^i_{n'\nu'}
=\tilde{\lambda}_i u^i_{n\nu},
\label{eigen_tilde_N}
\end{equation}
for each parity and angular momentum sector.
The existence of small eigenvalue, $\tilde{\lambda}_i \ll 1$,
indicates a strong overcompleteness of the basis set.
We impose the condition on the eigenvalues,
that the smallest eigenvalue, $\tilde{\lambda}_\text{min}$,
must be greater than $10^{-3}$.
This is done by the following procedure.
First, we calculate eigenvalues of $2 \times 2$ matrix of 
${\tilde N}^{J(\pm)}_{n\nu, n'\nu'}$ for all possible pairs of
$(\ket{\Psi_{n\nu}^{J(\pm)}}, \ket{\Psi_{n'\nu'}^{J(\pm)}})$. 
If the smaller eigenvalue is less than $10^{-3}$, we remove
one of them  according to
the magnitude of its diagonal element
(remove $\ket{\Psi_{n'\nu'}}$ if
${\tilde N}^{J(\pm)}_{n'\nu', n'\nu'} < {\tilde N}^{J(\pm)}_{n\nu, n\nu}$).
The number of basis states, $\{ \ket{\Psi_{n\nu}^{J(\pm)}}\}$
surviving these screenings
is now denoted as $N_{\rm sc}$. 
For the $J^{\pi}=3^-$ states of $^{16}$O,
$N_\text{sc}$ is of order of one hundred.
If the $\tilde{\lambda}_\text{min}$ of the 
$N_\text{sc}\times N_\text{sc}$ matrix,
${\tilde N}^{J(\pm)}_{n\nu, n'\nu'}$,
is larger than $10^{-3}$.
we can proceed to the configuration mixing calculation
to solve Eq.~(\ref{eq:HWEQ}).
Otherwise, we will further reduce the number of states:
We diagonalize the matrix ${\tilde N}^{J(\pm)}_{n\nu, n'\nu'}$ in a space
spanned by the basis except for a single state,
$\ket{\Psi_{m\mu}^{J(\pm)}}$.
This is the diagonalization of the $(N_{\rm sc}-1) \times (N_{\rm sc}-1)$ 
matrix.
We do this $N_\text{sc}$ times for all possible $\ket{\Psi_{m\mu}^{J(\pm)}}$,
in order to find the one, $\ket{\Psi_{m\mu}^{J(\pm)}}_\text{ex}$, for which
the minimum eigenvalue of the remaining
$(N_{\rm sc}-1) \times (N_{\rm sc}-1)$ matrix becomes the largest.
This state, $\ket{\Psi_{m\mu}^{J(\pm)}}_\text{ex}$,
is removed from the basis set.
This process is repeated and the number of basis is reduced
one by one, until all the eigenvalues,
${\tilde\lambda}_i$ ($i=1,\cdots,N_\text{b}^{J(\pm)}$),
becomes larger than $10^{-3}$.
In the case of $J^{\pi}=3^-$ in $^{16}$O,
several dozens of configurations of $\ket{\Psi_{m\mu}^{3(-)}}$
are discarded in this screening,
so that the final number of the basis states is $N_\text{b}^{3(-)}\approx 50$.
Note that $N_\text{b}^{J(\pm)}$ is the number of states in
$\{\ket{\Psi_{n\nu}^{J(\pm)}}\}$,
thus the number of adopted S-det's (that of $\ket{\Phi_n}$)
is in general less than $N_\text{b}^{J(\pm)}$.

In order to check numerical accuracy and stability,
it is convenient to define
a ``normalized'' norm eigenstate corresponding to
an eigenvalue $\tilde\lambda_i$ as
\begin{equation}
\ket{\Psi^{JM(\pm)}_i}_M \equiv
\frac{1}{\sqrt{\tilde{\lambda}_i}}
\sum_{n\nu}
\frac{u^i_{n\nu}}{\sqrt{N^{J(\pm)}_{n\nu,n\nu}}}
\sum_{K}v^{n\nu}_{K}P^{J}_{MK}P^{\pm}\ket{\Phi_n} .
\label{eq:norm_eigenstate}
\end{equation}
Using these states as a basis,
we calculate the norm and Hamiltonian kernel matrices,
\begin{equation}
\tilde{N}^{J(\pm)}_{ij} \equiv \inproduct{\Psi_i^{J(\pm)}}{\Psi_j^{J(\pm)}} ,
\quad\quad
\tilde{H}^{J(\pm)}_{ij} \equiv \bra{\Psi_i^{J(\pm)}} H \ket{\Psi_j^{J(\pm)}} ,
\end{equation}
and solve the generalized eigenvalue equation
\begin{eqnarray}
\sum_{j=1}^{N_\text{b}^{J(\pm)}}\left\{{\tilde H}^{J(\pm)}_{ij}
-E^{J(\pm)}{\tilde N}^{J(\pm)}_{ij}\right\}\tilde{g}_j^{J(\pm)}=0.
\label{eq:HWEQ}
\end{eqnarray}
We obtain the energy eigenvalues $E^{J(\pm)}$ and the eigenvectors
$\tilde{g}_i^{J(\pm)}$.

\begin{figure*}[t]
\begin{minipage}{0.4\textwidth}
\includegraphics[width=\textwidth]{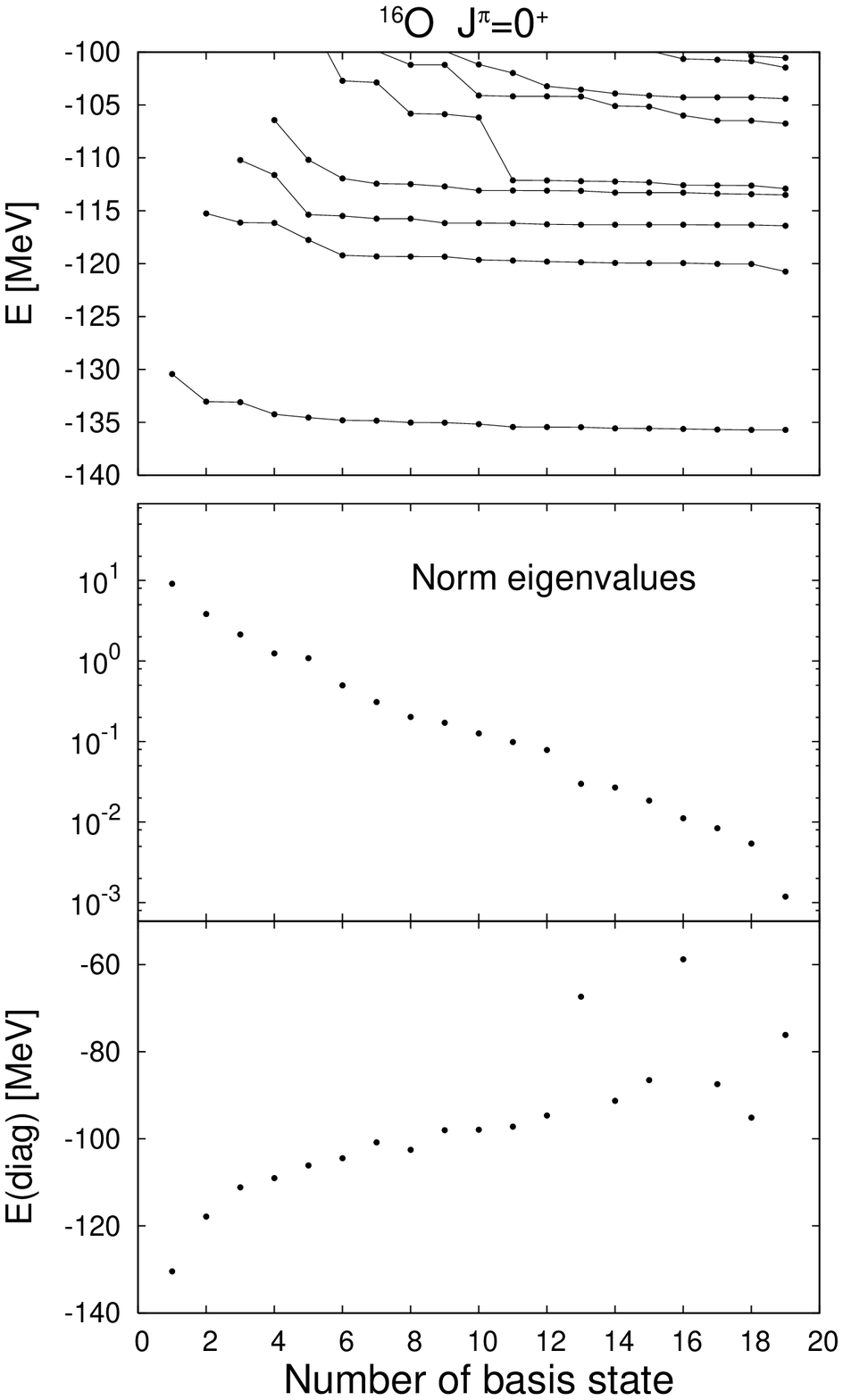}
\end{minipage}
\begin{minipage}{0.4\textwidth}
\includegraphics[width=\textwidth]{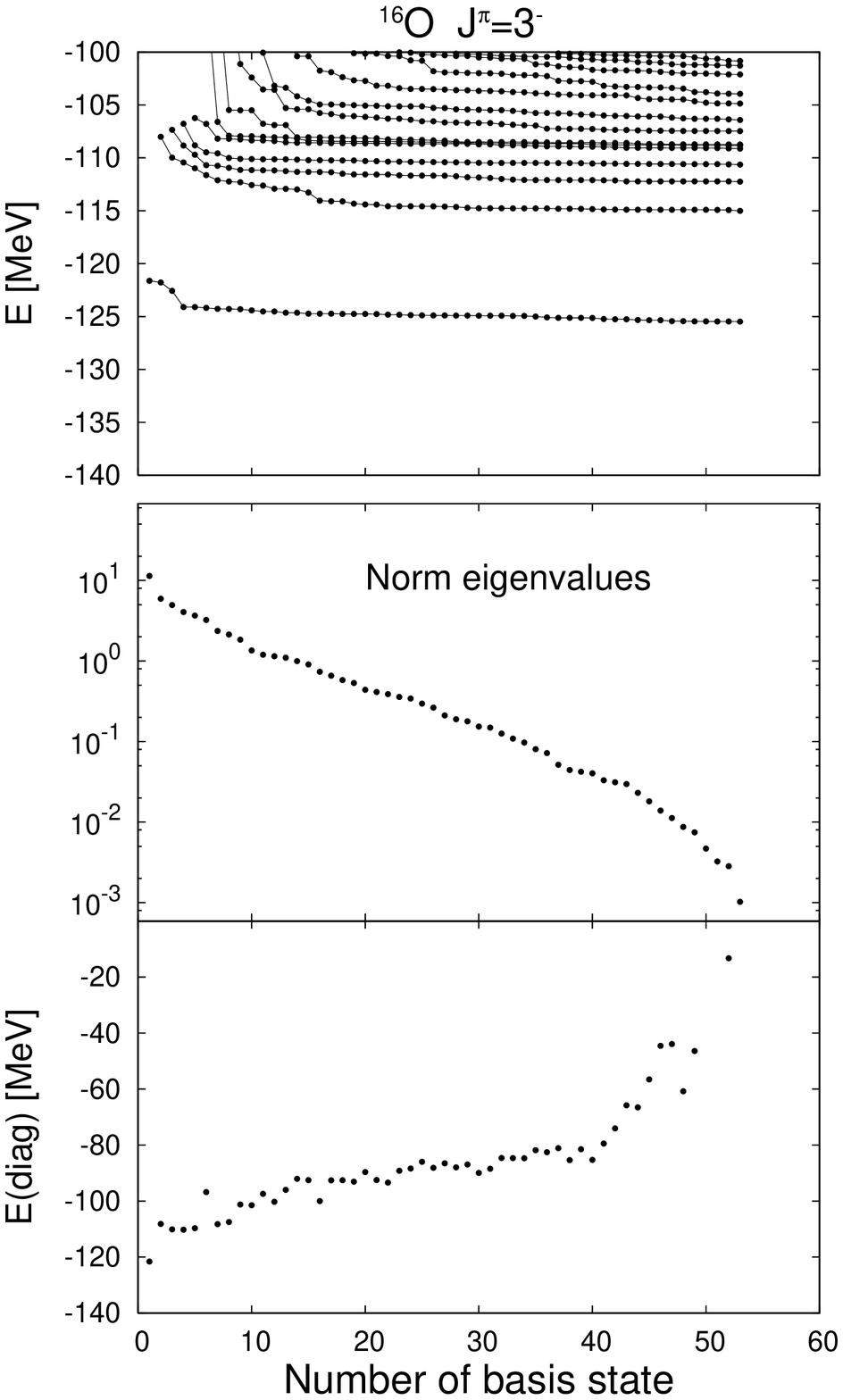}
\end{minipage}
\caption{\label{fig:norm}
Energies and norm eigenstates of $J^{\pi}$=$0^{+}$ states (left)
and $3^{-}$ states (right) in $^{16}{\rm O}$.
The top panels show calculated energy $E^{J(\pm)}$ of Eq.~(\ref{eq:HWEQ})
as a function of dimension of the adopted model space.
The middle and bottom panels show $\tilde\lambda_i$ and
$\tilde{E}_i^{J(\pm)}=\tilde{H}_{ii}^{J(\pm)}$, respectively.
See text for details.
}
\end{figure*}

\subsection{Quality of solutions}

In this section, we examine quality of
solutions obtained by diagonalizing
Eq.~(\ref{eq:HWEQ}) and how ``{\it complete}'' the selected basis is.
Let us emphasize again that we do not intend to obtain the exact eigenstates
of a given Hamiltonian.
The exact ground state of a Hamiltonian with the zero-range interaction
such as Eq.~(\ref{BKN}) perhaps leads to an unphysical solution.
Instead, we aim to take into account correlations of
its long-range part only.
Therefore, we examine whether the method can produce convergent results
for low-lying states.

Let us suppose that we have selected
$N_\text{b}^{J(\pm)}$ basis states for specific parity and angular momentum,
$(J,M,\pm)$.
We solve Eq.~(\ref{eigen_tilde_N}) to obtain eigenvalues,
$\tilde{\lambda}_i$, and vectors, $u_{n\nu}^i$, then
construct the norm eigenstates of Eq.~(\ref{eq:norm_eigenstate}),
$\ket{\Psi_i^{J(\pm)}}$.
First, the states $\{\ket{\Psi_i^{J(\pm)}}\}$ are sorted
according to the magnitude of their norm eigenvalues,
$\tilde{\lambda}_i$.
Thus, the eigenstates
are arranged in sequence of
$\tilde{\lambda}_1>\tilde{\lambda}_2
 >\cdots >\tilde{\lambda}_{N_\text{b}}$.
The middle and bottom panels in Fig.~\ref{fig:norm}
show distributions of $\tilde{\lambda}_i$ and
the diagonal elements of the Hamiltonian,
$\tilde{E}_i^{J(\pm)}=\tilde{H}_{ii}^{J(\pm)}$, respectively,
for $J^\pi=0^+$ (left) and $3^-$ states (right) in $^{16}$O.
In this calculation,
there are 19 basis states for $J^\pi=0^+$ and 53 for $J^\pi=3^-$
for which all the norm eigenvalues are larger than $10^{-3}$.
The $\tilde\lambda_i$ decrease linearly in the logarithmic scale.
An interesting thing is the fact that
the energy expectation values, $\tilde{E}_i^{J(\pm)}$,
are closely correlated with
the norm eigenvalues $\tilde\lambda_i$.
The energies $\tilde{E}_i^{J(\pm)}$ roughly show a monotonic increase with $i$,
as $\tilde\lambda_i$ decrease.
This may justify the screening process to discard
states with small norm eigenvalues,
because those states possess large energy expectation values
and are expected not to play a significant role for
low-energy excitations.

The top panels in Fig.~\ref{fig:norm} show
resultant energy eigenvalues, $E^{J(\pm)}$, obtained
by the diagonalization of Eq.~(\ref{eq:HWEQ}).
The horizontal axis indicates the number of basis states included in
the calculation, which is increased one by one from left to right;
$\{ \ket{\Psi_1^{J(\pm)}} \}$,
$\{ \ket{\Psi_1^{J(\pm)}},\ket{\Psi_2^{J(\pm)}} \}$, $\cdots$,
$\{ \ket{\Psi_1^{J(\pm)}},\cdots, \ket{\Psi_{N_\text{b}}^{J(\pm)}} \}$.
Calculated spectra for $J^\pi=0^+$ and $3^-$ states in $^{16}$O
are shown in the left and right panels, respectively.
As is seen in the figure,
low-energy spectra become almost invariant with respect to the
inclusion of new basis states.
In other words,
energies of the ground and low-lying states are insensitive
to the inclusion of states with small norm eigenvalues.
These convergent behaviors suggest that the long-range correlations
for low-lying states are taken into account in the calculation.

\begin{figure*}[tb]
\includegraphics[width=\textwidth]{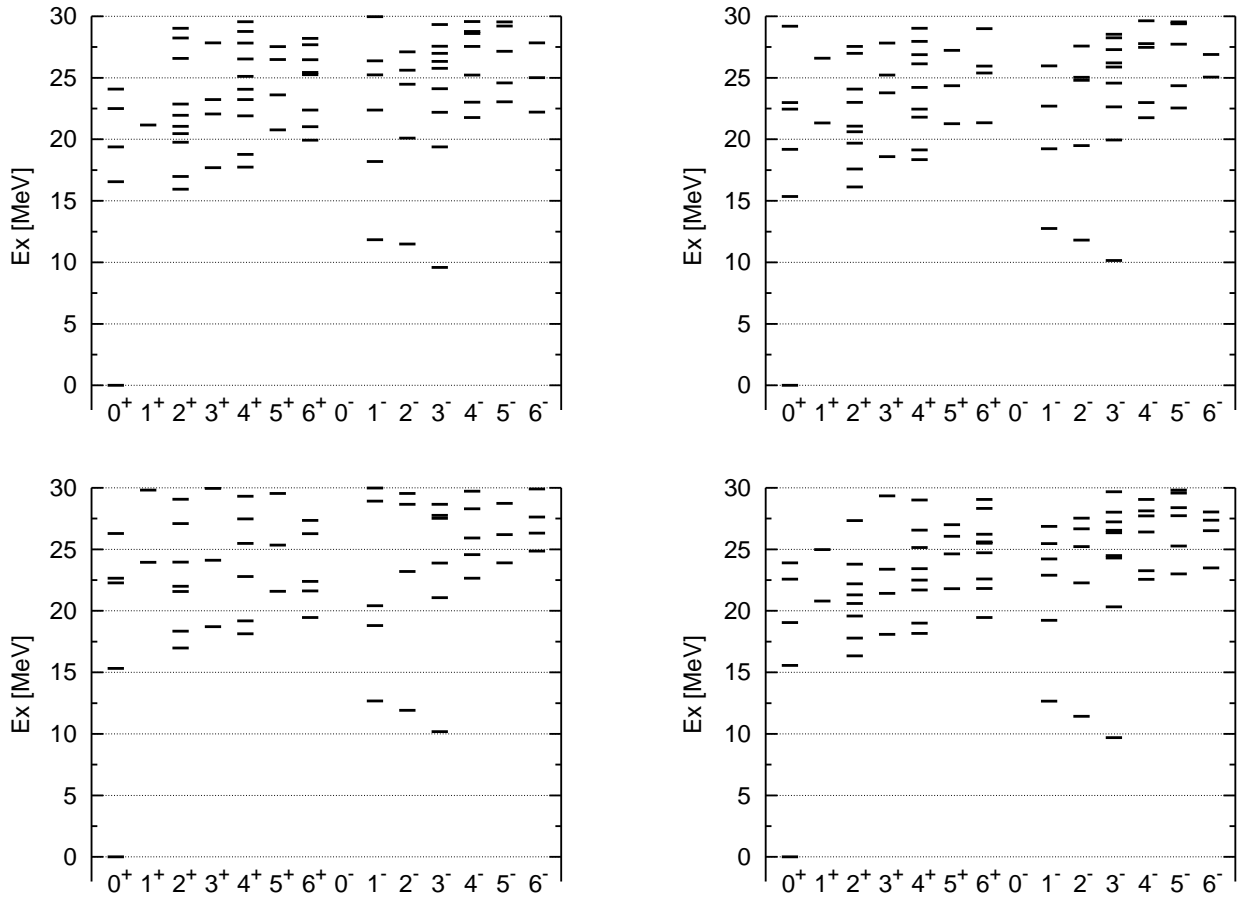}
\caption{\label{fig:compare}Excitation energies of $^{16}$O.
Symbols at the bottom of each panel indicate the quantum numbers, $J^\pi$.
Results of calculations employing four different sets of Slater 
determinants are displayed.
}
\end{figure*}

To further examine the completeness of the basis states,
we check the identity of results produced with
different sets of basis states
(initially generated with different random numbers).
If our prescription provides a complete set of basis
for the long-range correlations of the Hamiltonian,
the energy spectra should not depend on the initial S-det's
from which the imaginary-time iteration started.
In Fig.~\ref{fig:compare}, excitation energies in the $^{16}$O 
nucleus calculated with four different sets of S-det's are compared.
In these four independent calculations,
different seeds for the random number
were used in preparing the initial S-det's.
The energies of the lowest and the next lowest states
for each parity and angular momentum $(J^\pi)$ coincide
to each other in reasonable accuracy.
For example,
negative-parity excitations of $1^-$, $2^-$, and $3^-$
states appear below 15 MeV, and there are no other states
in this energy region.
The results become less reliable for higher states in each $J^\pi$ sector.
The second $0^+$ state (first excited $0^+$)
appears around $15\sim 17$ MeV in all calculations.
However, excitation energy of the third (second excited) $0^+$ state
in the bottom-left panel is notably higher than the other three.

The excitation energy of the second $0^+$ state
is much higher than the experimental value (6.05 MeV).
The three negative-parity
excited states with $J^\pi=3^-$, $1^-$, and $2^-$,
are observed at 6.13, 7.12, and 8.87 MeV, respectively.
Since the BKN interaction adopted in the present work,
which does not contain the spin-orbit interaction,
is too simple to give a quantitative description of nuclear structure,
one should not seriously take these discrepancies
between the calculation and the experiment.

We also perform the same examination for other nuclei discussed below,
$^{12}$C and $^{20}$Ne.
The final number of basis states $N_\text{b}^{J(\pm)}$
and behavior of the convergence is similar to those of $^{16}$O.
The comparison of results among sets of basis generated
with different random numbers
may provide a useful information about reliability of calculations.
From these analysis, we may judge
how many eigenstates in each $J^\pi$ sector can be trusted.

Before finishing this section,
let us comment on the cut-off value of the norm eigenvalues.
In the bottom panels of Fig.~\ref{fig:norm},
the energy expectation values become
somewhat scattered as the norm eigenvalues, $\tilde\lambda_i$,
approach to $10^{-3}$.
This may be an indication of the numerical instability.
In fact, if we include basis functions
with $\tilde\lambda_i<10^{-3}$,
the configuration mixing leads to unphysical solutions.
For instance, if we set a cut-off of $\tilde\lambda_i > 10^{-4}$,
the ground state becomes completely different
from the Hartree-Fock state
and its energy is unreasonably lowered by the diagonalization
of the Hamiltonian.
It is most likely that
this problem originates from the numerical error in 
evaluating the matrix elements, especially
in the angular momentum projection.

\section{\label{sec:C_Ne}Energy spectra of $^{12}$C and $^{20}$N\lowercase{e}}

\begin{figure}[tb]
\begin{minipage}{0.4\textwidth}
\includegraphics[width=\textwidth]{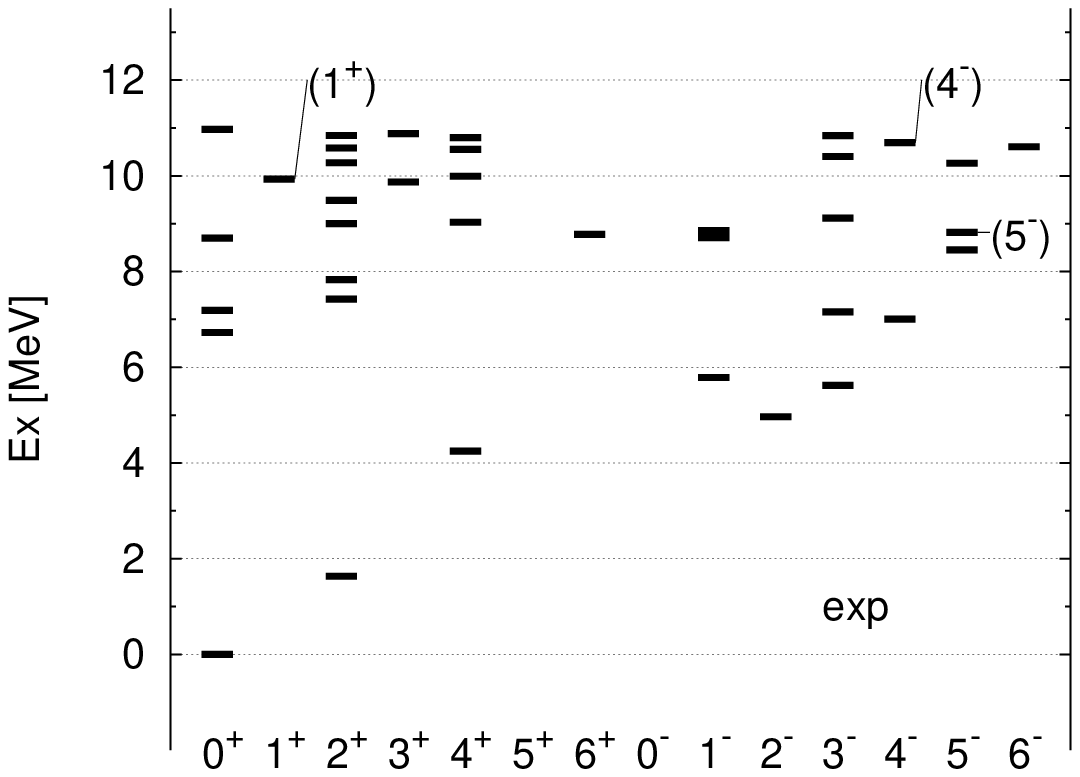}
\end{minipage}
\begin{minipage}{0.4\textwidth}
\includegraphics[width=\textwidth]{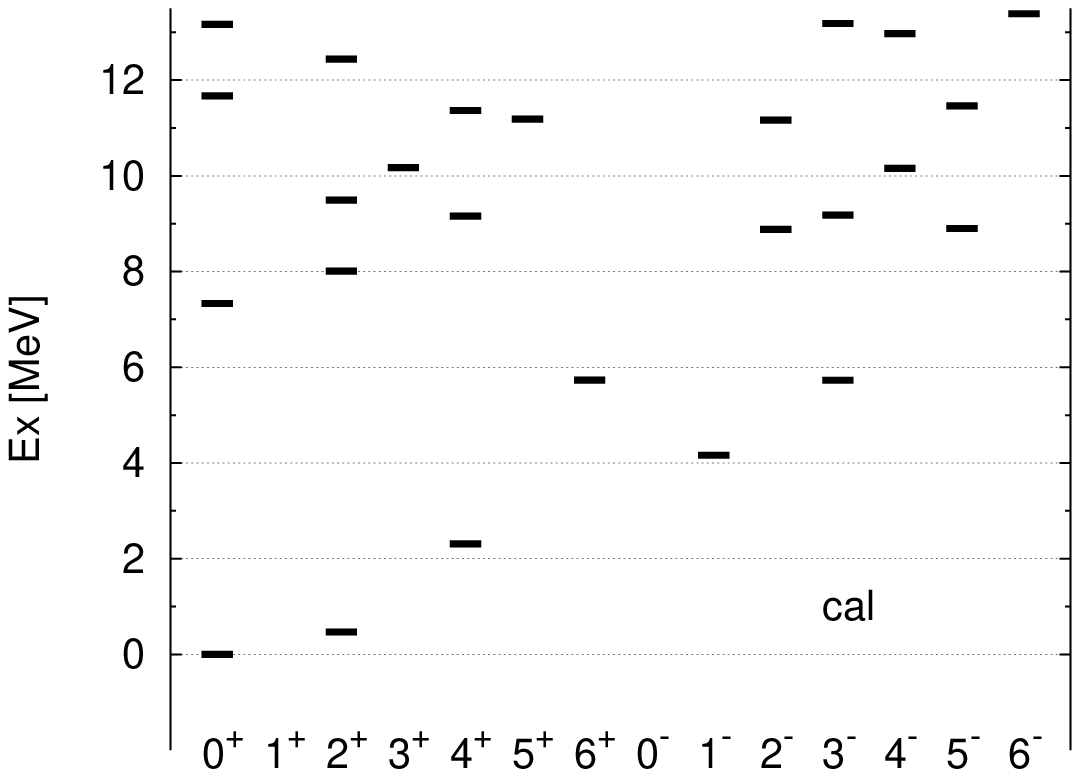}
\end{minipage}
\caption{\label{fig:20Ne}
Calculated and experimental excitation energy spectra of $^{20}{\rm Ne}$.
Symbols at the bottom of each panel indicate the quantum numbers, $J^\pi$.
}
\end{figure}

\begin{figure}[tb]
\begin{minipage}{0.4\textwidth}
\includegraphics[width=\textwidth]{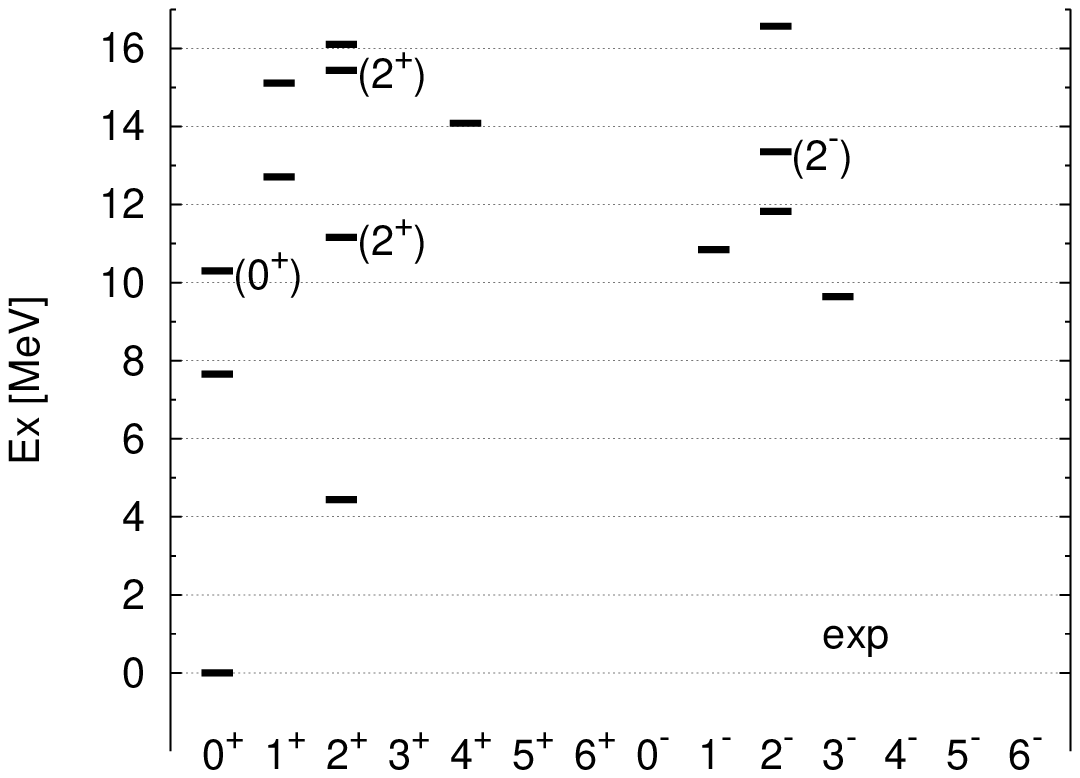}
\end{minipage}
\begin{minipage}{0.4\textwidth}
\includegraphics[width=\textwidth]{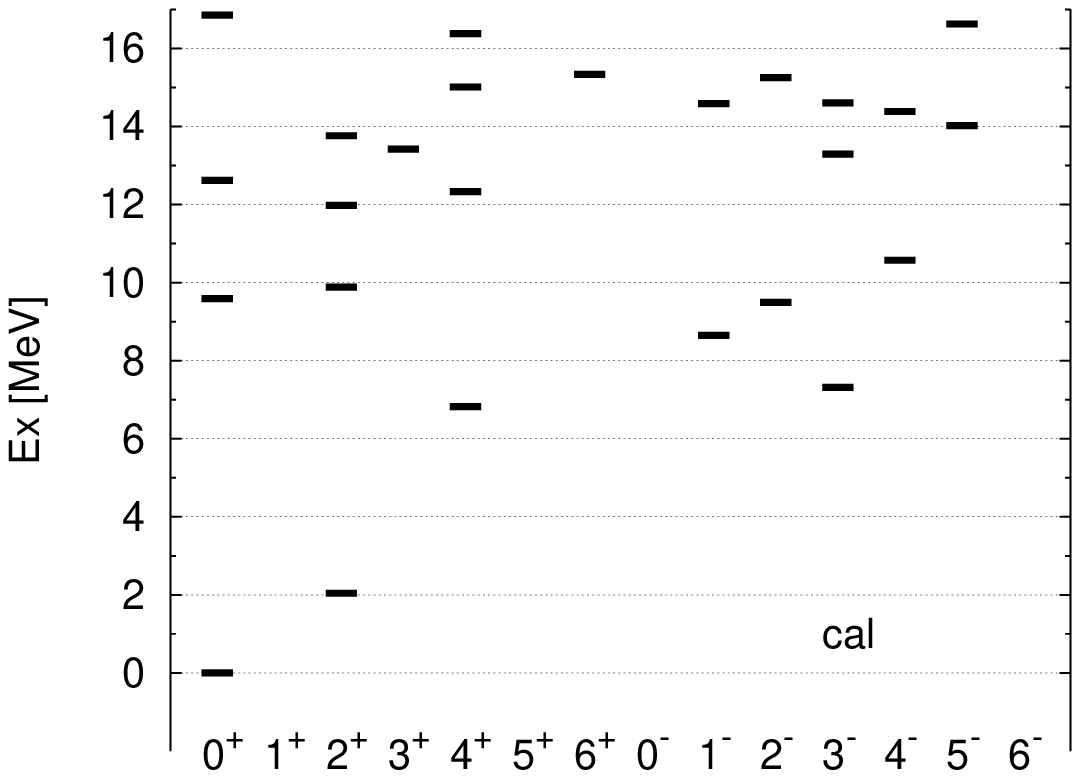}
\end{minipage}
\caption{\label{fig:12C}
The same as Fig.~\ref{fig:20Ne} but for $^{12}{\rm C}$.
}
\end{figure}

The BKN interaction of Eq.~(\ref{BKN}) is adopted for
testing our new method.
As we have mentioned before, one should not expect
a quantitative description of the low-lying spectra.
However, the present calculation gives
a reasonable description for some excited states of light nuclei,
especially for those composed of the $LS$-closed clusters.
In this section, we present
calculated energy spectra of $^{20}$Ne and $^{12}$C nuclei.
In these nuclei, there appear
the $LS$-closed clusters in the ground and excited states
($\alpha$+$^{16}$O for $^{20}$Ne and 3$\alpha$ for $^{12}$C).
In this paper, we restrict our discussion on the energy spectra
in these $N=Z$ even-even nuclei.
A detailed discussion on the structure of excited states including 
information on the transition matrix elements will be given 
in our next work employing a realistic Skyrme interaction. 

Figure~\ref{fig:20Ne} shows calculated energy spectra (left panel)
and those in measurement (right) in $^{20}$Ne.
It has been well-known that the 
$K^{\pi}=0^+$ ground state band and the $K^{\pi}=0^-$ 
negative-parity band starting with $1^-$ state at 5.785 MeV 
constitute a kind of inversion doublet band of the $\alpha$-$^{16}$O 
cluster structure.
This inversion doublet bands are
reasonably described in our calculation.
In the measurement, 
the lowest negative parity band is the $K^{\pi}=2^-$ band 
at 4.968 MeV.
In our calculation, it is around 9 MeV,
reflecting the importance of the spin-orbit splitting
of $p$ and $d$ orbitals for this excitation.

We next discuss results for $^{12}$C whose spectra are shown in
Fig.~\ref{fig:12C}. Again, the calculation (left panel) is
compared with measured spectra (right). The calculation 
produce the ground state rotational band,
but its moment of inertia is significantly larger than
the observed values.
In the negative parity, our calculation produces $3^-$ state 
in the lowest energy, also $1^-$ and $2^-$ states at low excitation energies.
These are qualitatively in agreement with experiments.
In the positive parity excited states, 
the calculation indicates two $0^+$ states around 10 and 12 MeV.
These may correspond to the measured states around 8 and 10 MeV. 

In Refs. \cite{PPHF},
we have reported the variation after parity projection
calculation employing the BKN interaction.
The angular
momentum projection after variation is achieved in the
calculation employing Skyrme force \cite{PPSHF,PhD:Ohta,PPSHF-P}.
A part of results presented
in this section coincide fairly well with these variation after
parity projection calculations.
This suggests
that the variation after parity projection gives a dominant correlation
for a certain class of states (with clustering)
in $^{12}$C and $^{20}$Ne.

\section{\label{sec:summary}Summary}

In this paper, we report our attempt to develop a new
computational method to include all the long-range correlations
beyond the mean-field approximation.
We aim at a systematic description of the ground and low-lying
excited states using a mean-field Hamiltonian,
without assuming their structure {\it a priori}.

First, we stochastically generate many Slater determinants.
The single-particle orbitals are expressed on the three-dimensional
Cartesian grid representation.
In order to remove high-energy components
in those states, we use the imaginary-time iteration method.
The imaginary-time evolution produces many trajectories important for
low-energy modes of excitation.
We select some of these states to keep the linear independence.
We then project them on
good parity and angular momentum, and perform configuration 
mixing calculation.
The BKN interaction is utilized to examine feasibility and
difficulty of the method.
We have found that there is a numerical difficulty to achieve 
the configuration mixing calculation.
The eigenvalues of the norm matrix can
be close to zero, when the selected states are overcomplete.
In the practical calculations,
a small numerical error in the angular momentum projection results in
the occurrence of negative eigenvalues of the norm matrix.
We eliminate states responsible for these zero and negative eigenvalues
before solving the generalized eigenvalue problem.

We show calculated results for some light $4N$ nuclei,
$^{12}$C, $^{16}$O, and $^{20}$Ne. In these nuclei, appearances of
various cluster states are known in excited states. 
Our calculation provides reasonable excitation energies
for  $\alpha + ^{16}$O states of $^{20}$Ne and $3\alpha$ states 
of $^{12}$C for which the spin-orbit interaction, which is
not included in the BKN force, does not play an important role.

The results calculated with different sets of random numbers
are approximately identical to each other.
However, the discrepancy becomes more evident for states at higher
energies.
In the present level of accuracy,
we may predict the lowest and possibly the second lowest states in
each parity and angular momentum sector.
Improvement in numerical accuracy,
especially in the three-dimensional angular momentum projection,
is desired for future work with more realistic interactions.

\begin{acknowledgments}
This work is supported in part by the Grant-in-Aid for Scientific Research
in Japan (Nos. 17540231 and 18540366).
We thank the Yukawa Institute for Theoretical Physics (YITP)
at Kyoto University and the Institute for Nuclear Theory:
Discussions during the workshops on
``New Developments in Nuclear Self-Consistent Mean-Field Theories''
(YITP-W-05-01)
and those during ``Nuclear Structure Near the Limits of Stability'' (INT-05-3)
were useful to complete this work.
The numerical calculations
have been performed at SIPC, University of Tsukuba,
at RCNP, Osaka University.  and at YITP, Kyoto University.

\end{acknowledgments}

\appendix*
\section{Matrix elements between non-orthogonal Slater determinants}
\label{sec:matrix_elements}

In this appendix, we present useful expressions for calculating
matrix elements such as the integrand of Eq.~(\ref{eq:ME}).
In general, we discuss transition amplitude of an operator $\hat{O}$
between two different S-det's, $\bra{\Phi} \hat O \ket{\Psi}$.
The S-det's are expressed
in terms of orthonormal single-particle orbitals,
$\ket{\Phi}=\det \{ \ket{\phi_i(j)} \}/\sqrt{A!}$ and
$\ket{\Psi}=\det \{ \ket{\psi_i(j)} \}/\sqrt{A!}$,
with $\inproduct{\phi_i}{\phi_j}=\inproduct{\psi_i}{\psi_j}=\delta_{ij}$.
Here, we assume $\ket{\Phi}$ and $\ket{\Psi}$
are not orthogonal to each other.
To calculate these matrix elements, it is
convenient to define following orbitals,
\begin{equation}
\ket{\tilde \psi_i}
= \sum_j \ket{\psi_j} (B^{-1} )_{ji}
\end{equation}
where the matrix $B$ is defined by 
$B_{ij}=\inproduct{\phi_i}{\psi_j}$.
The overlap matrix element is given by the determinant of $B$,
$\inproduct{\Phi}{\Psi}=\det B$.
It can be easily confirmed that
$\ket{\tilde \psi_i}$ 
are bi-orthogonal to $\ket{\phi_j}$,
having
\begin{equation}
\label{bi-orth}
\inproduct{\phi_i}{\tilde\psi_j} = \delta_{ij}.
\end{equation}
We also note that the S-det constructed from 
$\ket{\tilde\psi_i}$ is proportional to $\ket{\Psi}$,
\begin{equation}
\ket{\tilde\Psi}\equiv
 \frac{1}{\sqrt{A!}} \det \left\{ \tilde\psi_i(j) \right\}
= \frac{\ket{\Psi}}{\inproduct{\Phi}{\Psi}} .
\end{equation}
Since the S-det $\ket{\Psi}$ is now represented by
single-particle orbitals $\ket{\tilde\psi_i}$ which
have a bi-orthogonal property (\ref{bi-orth}),
the matrix elements $\bra{\Phi} \hat O \ket{\Psi}$ can be
expressed in a familiar form very similar to that of the expectation value,
$\bra{\Phi} \hat O \ket{\Phi}$.
Suppose that, the expectation value of the operator $\hat{O}$,
$\langle \Phi \vert \hat O \vert \Phi \rangle$,
is expressed as a functional of the density matrix,
$\rho_{ij}(\Phi)=\bra{\Phi} c_j^\dagger c_i \ket{\Phi}$,
\begin{equation}
 O[\rho(\Phi)] = \bra{\Phi} \hat O \ket{\Phi}.
\end{equation}
Then, the matrix element between $\ket{\Phi}$ and $\ket{\Psi}$
is expressed as
\begin{equation}
\bra{\Phi} \hat O \ket{\Psi}
= \bra{\Phi} \hat O \ket{\tilde\Psi} \inproduct{\Phi}{\Psi}
= O[\tilde \rho(\Phi\Psi)]  \inproduct{\Phi}{\Psi} ,
\end{equation}
where the interstate density $\tilde\rho(\Phi\Psi)$ is defined by
\begin{equation}
\label{tilde_rho}
\tilde \rho_{ij}(\Phi\Psi) = \bra{\Phi} c_j^\dagger c_i \ket{\tilde\Psi} .
\end{equation}
Therefore, the matrix element between two S-det's,
$\bra{\Phi} \hat O \ket{\Psi}$, is given by $O[\rho]\times \det B$,
where the density matrix $\rho$ is replaced by the
interstate density matrix (\ref{tilde_rho}).

\end{document}